\documentstyle[11pt,
epsfig]{article}
\input{psfig.sty}
\newcommand{\lx}{\lambda}
\newcommand{\Lx}{\Lambda}

\newcommand{\rht}{\tilde{\rho}}

\newcommand{\be}{\begin{equation}}
\newcommand{\ee}{\end{equation}}
\newcommand{\een}{\end{subequations}}
\newcommand{\ben}{\begin{subequations}}
\newcommand{\beq}{\begin{eqnarray}}
\newcommand{\eeq}{\end{eqnarray}}

\def \lta {\mathrel{\vcenter
     {\hbox{$<$}\nointerlineskip\hbox{$\sim$}}}}
\def \gta {\mathrel{\vcenter
     {\hbox{$>$}\nointerlineskip\hbox{$\sim$}}}}

\def\MPL#1{Mod. Phys. Lett.~{\bf #1}}
\def\NP#1{Nucl. Phys.~{\bf #1}}

\def\PL#1{Phys. Lett.~{\bf #1}}
\def\PR#1{Phys. Rev.~{\bf #1}}
\def\PRP#1{Phys. Rep.~{\bf #1}}

\def\ZP#1{Z. Phys.~{\bf #1}}

\begin{document}
\centerline{~ \hfill     UA/NPPS-12-2000}
\centerline{\phantom{A}}
\vspace{3.cm}
\centerline{\Large\bf
Quantum-Mechanical Tunnelling and the Renormalization Group
}
\vspace{2.5cm}
\centerline{{\large\bf A.S. Kapoyannis$^{a}$} {\large and}
{\large\bf N. Tetradis$^{a,b}$}}\vspace{0.3cm}
\centerline{\em a) Nuclear and Particle Physics Sector}
\centerline{\em University of Athens}
\centerline{\em GR-15771 Athens, Hellas}
\vspace{0.2cm}
\centerline{\em b) Department of Physics}
\centerline{\em University of Crete}
\centerline{\em GR-71003 Heraklion, Greece}
\vspace{4.cm}
\centerline{\large\bf Abstract}
\begin{quote}\large\indent
We explore the applicability of the exact renormalization group 
to the study of tunnelling phenomena. We investigate quantum-mechanical
systems whose energy eigenstates are affected significantly
by tunnelling through a barrier in the potential. 
Within the approximation of the
derivative expansion, we find that the exact renormalization group  
predicts the correct qualitative behaviour for the lowest
energy eigenvalues. However, quantitative accuracy is achieved only 
for potentials with small barriers. For large barriers, the use
of alternative methods, such as saddle-point expansions,
can provide quantitative accuracy. 
\end{quote}
\vspace{1cm}

\newpage

\paragraph{Introduction:} 
The exact renormalization group \cite{kad,wil}
is a powerful method with a wide range of applications in various fields. 
It provides a framework in which it is possible to 
study non-perturbative aspects of physical phenomena. 
This is made possible by an exact renormalization-group equation  
\cite{wil,rg,wet}
that describes the dependence of generating functionals for the
correlation functions of the theory on a coarse-graining scale.
In particular formulations, such as the effective average action
\cite{averact,review}, an exact equation can be obtained for
the flow of the coarse-grained free energy of the system, a quantity 
with intuitive physical interpretation. 

The biggest difficulty one faces in this approach concerns the
approximations that must be made in order to turn the exact
renormalization-group equation, a functional differential equation, 
into evolution equations for quantities such as the effective
potential. A widely used approximation method employs an
expansion of the effective action in the number of 
derivatives of the fields appearing in it. Its validity, at the
formal and practical level, has been studied and tested extensively
\cite{review}--\cite{more}. The absence of a small expansion 
parameter makes the precision of a given truncation in the number of
derivatives hard to
estimate, and one often must rely on comparisons with alternative methods.
However, one can obtain answers for quantities that are difficult to compute,
such as critical exponents, amplitudes and 
the critical equation of state for second-order phase transitions, or
the bubble-nucleation rate beyond 
the semiclassical level for first-order phase transitions \cite{review}.

In this letter we are interested in the applicability of the approach
to tunnelling phenomena. We work within the formulation of the
effective average action \cite{averact,review}. Recent work 
\cite{first1,first2} has demonstrated how the presence of a coarse-graining
scale leads to a consistent quantitative description of such phenomena, even
beyond the leading semiclassical level.
Of related interest is the question of the convexity of the effective
potential in field theory, 
which is induced by tunnelling configurations.
This issue has been addressed within the renormalization-group
approach as well \cite{convex}.

The study of the above problems within field theory is often obscured by
technical difficulties, such as the need to regulate the ultraviolet
divergences, or possible infrared problems arising from 
the presence of massless modes in the phase with spontaneous symmetry
breaking. A much simpler framework is provided by quantum
mechanics. One does not have to deal with ultraviolet problems and
spontaneous symmetry breaking does not arise. Moreover, the exact answers are
well known and comparisons are straightforward. A recent study 
\cite{qmt} has also explored the issues we address in this letter.
However, it employed an expansion in powers of
the field, an approximation that has a significant effect on the results. 
Our treatment goes beyond this approximation and leads to a more complete
conceptual understanding.

\paragraph{The problem:}
We would like to understand how well quantum-mechanical tunnelling 
is described by the exact renormalization-group approach. 
This approach employs field-theoretical language. However, a 
theory of a real scale field in one time and zero space dimensions
can be viewed as a quantum-mechanical system. This mapping
is obvious in the functional integral formulation 
\cite{ramond} through the 
replacement of the field $\phi$ by the position $x$ of
the quantum-mechanical particle. 
As a result, approximations employed in the study of field theories
in more than one space-time dimensions can be checked through comparison with 
exact results in one dimension.

We employ the formalism of the effective average action $\Gamma_k$
\cite{averact,review} that can be interpreted as a coarse-grained free energy. 
All fluctuations of the system with characteristic momenta larger
than an infrared cutoff 
$k$ (and wavelengths less than $2 \pi/k$) are integrated
out and incorporated in the effective couplings appearing in
$\Gamma_k$. The dependence of $\Gamma_k$ on the coarse-graining
scale $k$ is described by an exact renormalization-group 
equation \cite{wet}. An expansion in derivatives
of $\phi$ results in evolution equations for the
functions multiplying the various terms in $\Gamma_k$.
These form an infinite system of coupled partial differential equations 
in $\phi$ and $k$. In practice one must truncate this system, 
by keeping only a finite number of terms in the action.

In one Euclidean space-time dimension, 
the lowest truncation level (often referred to as the local potential
approximation) uses an action of the form\footnote{
We use $\hbar=1$ throughout the paper.}
\be
\Gamma_k = \int_{-\infty}^{\infty} dt \left\{
\frac{1}{2} \left(\frac{d \phi}{dt} \right)^2 + U_k(\phi)
\right\}.
\label{action} \ee
At the next level the derivative term is multiplied by a 
non-trival wavefunction renormalization $Z_k(\phi)$, while higher levels
include more derivatives of the field.
For the approximation of eq. (\ref{action}), 
the exact-renormalization group equation for $\Gamma_k$ results in
the evolution equation \cite{review,qmt}
\be
\frac{\partial U_k(\phi)}{\partial k}=
-\frac{1}{2 \pi} \log \left(1+ \frac{1}{k^2}\frac{\partial^2 U_k}{\partial 
\phi^2} \right)
\label{eveq} \ee
for $U_k(\phi)$. This equation has been derived for a sharp infrared 
cutoff $k$ that acts like a $\theta$-function,
completely integrating out
fluctuations of $\phi$ with momenta $q\geq k$ and excluding
fluctuations with $q<k$. It also appears in other formulations of the
exact renormalization group in the sharp-cutoff limit
\cite{rg,morris,qmt}.

In ref. \cite{qmt} $U_k(\phi)$ is further
expanded in powers of $\phi$, and eq. (\ref{eveq}) is turned into
a system of ordinary differential equations for the coefficients of
the expansion. This infinite system is truncated at a finite level.
In ref. \cite{convex} it was shown that such a truncation
does not reproduce correctly the solution of eq. (\ref{eveq})
in the regime that is relevant for the convexity of the effective
potential. For this reason, we do not make this additional approximation
here, but solve numerically the full partial differential equation
(\ref{eveq}) instead. For this we employ numerical algorithms discussed
in ref. \cite{num}.

The boundary conditions for the solution of eq. (\ref{eveq}) are 
fixed at a scale $k=\Lx$ much larger than the physical scales of the
low-energy theory. At this ``microscopic'' level, the potential is
determined by the fundamental theory. The case of one dimension
is particular in this respect, because the contribution of the 
ultraviolet regime to the evolution of the potential is negligible.
This reflects the absence of ultraviolet divergences in quantum
mechanics. As a result, one can interpret the initial condition
$U_\Lx$ as the potential at the level where no fluctuations of the
system are taken into account, i.e. the classical potential $V$.
The solution of eq. (\ref{eveq}) in the limit $k\to 0$ incorporates the
effect of fluctuations at all scales, and $U_0$ becomes the effective
potential $U$. In quantum-mechanical terms, $U$ determines the
expectation value of the  energy of the system 
for a given expectation value of the position.

We consider $Z_2$-symmetric classical potentials of the form
\be
U_\Lx=V=-\frac{1}{2}m^2 \phi^2 + \lx \phi^4 + \frac{m^2}{16 \lx}.
\label{initial} \ee
Their minima satisfy $U(\phi_{\rm min})=0$.
By expressing all dimensionful quantities in units of $m$, we
can set $m=1$ in the above relation. 
The presence of a barrier implies that tunnelling configurations
can be relevant for this system. 
For $\lx\to 0$ the height of the barrier becomes much larger
than the distance between the minima and the origin, 
and we expect tunnelling to play an important role.

\paragraph{The solution:}
Starting with the above initial condition, we solve the partial 
differential equation (\ref{eveq}) numerically. A typical solution
is presented in fig. 1 for $\lx=0.05$.
The evolution starts at $k=\Lx \gg 1$ with $U_\Lx=V$. It is apparent that
for $k\gg 1$ the potential $U_k$ changes very little. This means that the
ultraviolet (short-distance) fluctuations do not contribute to the
evolution, and reflects the absence of ultraviolet divergences in
quantum mechanics. For $k = {\cal O} (1)$ the evolution becomes fast, while
the location of the minima moves to zero. Eventually, the evolution slows down
for $k\to 0$ and the potential takes its final form with only one minimum
at the origin.

Several properties of this solution have physical significance:
\begin{itemize}
\item 
In one dimension,
the location of the minimum always runs to zero for $k \to 0$.
This reflects the absence of spontaneous symmetry breaking 
in quantum mechanics.
\item
The value of the potential at the minimum changes from zero to
a positive value. In more than one dimensions, the absolute scale
of the potential is related to the cosmological constant and is
very sensitive to the contribution from the ultraviolet fluctuations.
In one dimension, the ultraviolet contributions are negligible and
the absolute scale of the potential is meaningful. In fact, the
value of $U_0=U$ at the minimum corresponds to the expectation value
of the system in the vacuum, i.e. the eignevalue $E_0$ of the ground state.
In our case $E_0=U(0)$.
\item
In field-theoretical terms, the first excitation of the system above the
vacuum 
corresponds to a particle at rest. Its mass term is given by the second
derivative of the potential with respect to $\phi$ at the minimum. 
This means that the eigenvalue of the first excited quantum-mechanical
state can be computed as
$E_1=E_0 + \sqrt{\partial^2 U(0)/\partial \phi^2}$.
\item
If the lowest energy eigenvalue $E_0$ is below the maximum of the barrier of
the classical potential $V(0)=1/(16 \lx)$ (as in our example), 
tunnelling plays an important
role in the problem. The solutions are then of trully 
non-perturbative nature. 
\end{itemize}

In fig. 2 we plot $\left[\partial^2 U_k(\phi)/\partial \phi^2\right]/k^2$
during the whole evolution. The continuous lines depict the initial part
of the running with $\Lx\geq k\gta 1$, 
while the dashed ones the final part with $1\gta k \geq 0$.
We observe that there is a range of $k$ for which
$\partial^2 U_k(\phi)/\partial \phi^2 \simeq -k^2$ near the origin of the
potential. The reason for this behaviour 
is the presence of a pole at $-1$ 
for $\left[\partial^2 U_k(\phi)/\partial \phi^2\right]/k^2$ in eq.
(\ref{eveq}). As this pole cannot be crossed, 
$\partial^2 U_k(\phi)/\partial \phi^2$ remains close to $-k^2$ while $k$ is
reduced towards zero. The pole is approached only for $\lx \lta 0.1$ and the
part of the evolution the system spends near it
grows for $\lx \to 0$.
In this limit the importance of the barrier increases. 

This behaviour has been observed in studies of field theories with
spontaneous symmetry breaking in higher dimensions \cite{convex}.
It results in the presence of a flat part of the potential between the
two minima for $k\to 0$. ($\partial^2 U_{k\to 0}(\phi)/\partial \phi^2$
becomes zero between the minima.) It cannot be reproduced through the
expansion of $U_k(\phi)$ in powers of $\phi$ employed in ref. \cite{qmt}. 
In the one-dimensional case we are studying, 
the minimum of the potential always moves to
the origin and $\partial^2 U_k(0)/\partial \phi^2$ becomes positive
at the end of the evolution.
However, the longer the part of the evolution the system spends near
the pole, the smaller the final positive 
value of $\partial^2 U_{k\to 0}(0)/\partial \phi^2$. As this value
determines the difference between the first two energy eigenvalues,
the presence of the pole in the evolution equation predicts a vanishing
$E_1 - E_0$ for $\lx \to 0$. This is in qualitative agreement with
the expectation from quantum mechanics.

The quantitative accuracy of the predictions can be checked through 
comparison with the results of the 
numerical solution of the Schr\"odinger equation with
a potential given by eq. (\ref{initial}).
In fig. 3 we plot the values of $E_0$, $E_1$ predicted by the solution
of the evolution equation, along with the exact results, as a function
of $\lx$. 
We also show the height of the barrier $V(0)=1/(16 \lx)$.
Comparison with $E_0$, $E_1$ determines how important tunnelling is
for the corresponding solutions.
In fig. 4 we give values for
the difference $\Delta E = E_1-E_0$. We also include the 
prediction of the dilute-gas instanton approximation \cite{coleman,qmt}
\be 
\Delta E = 2 \sqrt{\frac{2 \sqrt{2}}{\pi \lx}} \exp 
\left( 
- \frac{1}{3 \sqrt{2}\lx}
\right).
\label{dilute} \ee

For $\lx\gta 0.15$
the agreement between the values of $E_0$ and $E_1$
resulting from our analysis and the exact results is at the few-per-cent 
level. For $\lx={\cal O}(1)$ it becomes 
better than 1\%.
The deviations are in the same direction, so that our prediction for the 
difference $\Delta E$ agrees at a level better than 1\% with the exact value
for all $\lx\gta 0.15$.
We conclude that the lowest order of the derivative expansion 
(eq. (\ref{action})) leads to very good quantitative results when 
the potential does not have a large barrier. 
This conclusion is reaffirmed through the study of a potential without
a barrier. In fig.~5 we give the results for the unharmonic oscillator, 
corresponding to the potential of eq. (\ref{initial}) with 
a mass term $m^2=-1$.
All this is in very good agreement with ref. \cite{qmt}, which indicates
that the additional approximation of the potential by a polynomial,
employed there, is very good for large $\lx$.

For $\lx \lta 0.15$ the lowest energy eigenvalue is below the top of the
barrier of the classical potential. This means that  
tunnelling plays a significant role in the solution.
In the region $0.09 \lta \lx \lta 0.15$ our results have an
accuracy better than 10\%. We conclude that the renormalization group
accounts reasonably well for tunnelling effects in this region.
For $\lx \lta 0.09$ the energy eigenvalue of the first excited 
state is below the top of the barrier as well. Our results have
large deviations from the exact values and the renormalization-group 
approach fails to give a good quantitative picture. However,
the correct qualitative behaviour is still reproduced.
In fig. 4 we observe that the difference $\Delta E$ diminishes with
decreasing $\lx$. Excessive requirements in computer time, in order to
reproduce correctly the solution near the pole,
forbid the numerical integration of eq. (\ref{eveq}) for very small $\lx$.
Already for $\lx=0.04$ the numerical error becomes close to 1\%.
However, as we explain below, we expect that $\Delta E$ becomes 0 for 
$\lx \to 0$. An extrapolation of our results in this limit, as well as an
approximate analytical solution (see below), verifies
our expectation.
However, the correct quantitative dependence
of $\Delta E$ on $\lx$, given by the dilute-gas instanton approximation
(eq. (\ref{dilute})), is not reproduced.

\paragraph{Discussion:}
In order to understand better the nature of the solutions of the
partial differential equation (\ref{eveq}), it is useful to
consider some approximations that permit an analytical treatment.
For large $k$ the potential has a minimum $\phi_0(k)$ away from the
origin. The $k$-dependence of the minimum can be obtained by considering
the total $k$-derivative of the condition 
$\partial U_k/\partial \phi |_{\phi=\phi_0}=0$.
One finds 
\be
\frac{d\phi_0}{d k}
=-\frac{1}{U''_k(\phi_0)}
\frac{\partial U'_k(\phi_0)}{\partial k} 
= \frac{1}{2\pi}\frac{U'''_k(\phi_0)}{U''_k(\phi_0)}
\frac{1}{k^2} \left[ 1+ \frac{U''_k(\phi_0)}{k^2} \right]^{-1},
\label{dphiz} \ee
where primes denote derivatives with respect to $\phi$.
For small $\lx$ the potential $U_k(\phi)$ 
can be approximated by a quartic polynomial, 
as in eq. (\ref{initial}). 
Moreover, the coefficient of the quartic term has a weak dependence
on $k$, as can be verified through the numerical solution.
Within perturbation theory, one can explain the smallness of
the omitted corrections by the fact that they involve powers of $\lx$. 

We consider first the case in which
${U''_k(\phi_0)}/{k^2} \ll 1$ during the whole evolution. 
Within our appoximations, eq. 
(\ref{dphiz}) can be integrated easily with the result
\be
\phi^2_0(k)=\phi^2_0(\Lx) - \frac{3}{\pi} \frac{1}{k},
\label{phizk} \ee
where we have assumed $\Lx \gg 1$. For any initial $\phi_0(\Lx)$
there is a value $k_{cr}=3/(\pi \phi^2_0(\Lx))=12\lx/\pi$  
at which the minimum moves to the
origin. This is in agreement with our expectation that 
spontaneous symmetry breaking cannot appear in one dimension.
At scales below $k_{cr}$ the minimum remains at the origin, while
the mass term obeys the equation
\be
\frac{dU''_k(0)}{d k}
= -\frac{1}{2\pi}\frac{U^{(4)}_k(0)}{k^2}
\left[ 1+ \frac{U''_k(0)}{k^2} \right]^{-1}.
\label{vz} \ee
For positive $U^{(4)}_k(0)$,
the mass term grows as $k$ is further reduced.
The evolution stops for very small $k$, when $U''_k(0) \gg k^2$
(decoupling).

It is possible that 
${U''_k(\phi_0)}/{k^2} ={\cal O} (1)$
before the minimum of the potential moves to zero. In the approximation
of a quartic potential with a constant quartic coefficient 
$\lx \lta 0.1$, we have
${U''_k(\phi_0)}/{k^2} \simeq 1$ for $k_p\simeq 1>k_{cr}$. 
This complicates the analytical study of eq. (\ref{dphiz}).
Moreover, it raises the possibility that, as  
$k^2$ is reduced below $k^2_p$, it may become much smaller than $U''_k(\phi_0)$
and the evolution may stop with the system in the phase
with symmetry breaking. This scenario is never realized because of the
presence of a pole at $\partial^2 U_k/\partial \phi^2=-k^2$ 
in the evolution equation (\ref{eveq}).
This pole is approached first at the origin, where eq. (\ref{vz})
applies. The relevant behaviour was analysed in detail
in ref. \cite{convex} for field theories in
more than one dimensions, where symmetry breaking can occur. 
It was shown that the solution
approaches the limit $k\to 0$ with $U''_k(0) \simeq -k^2$. 
The curvature at the origin vanishes for $k=0$ and this leads to
the convexity of the effective potential. 
Polynomial approximations of the potential, such as the one
employed in ref. \cite{qmt}, cannot reproduce this behaviour and
lead to the appearance of singularities at non-zero values of $k$.

Our numerical solution, depicted in figs. 1 and 2, reproduces the correct 
behaviour for $\lx=0.05$. The inner part of the potential becomes
very flat as the pole is approached. At some stage the flatness approaches
the minima and prevents the decoupling behaviour that would
occur for $U''_k(\phi_0) \gg k^2$. Eventually the minimum moves
to the origin and the system settles in the symmetric phase.
Because of the induced flatness near the origin, the mass term 
$\partial^2 U_{k\to 0}(\phi)/\partial \phi^2$ that determines the
energy difference $E_1-E_0$ is very small. As this type of 
behaviour becomes more pronounced in the limit $\lx\to 0$, we
expect that $E_1-E_0$ becomes zero in this limit. 

The unsatisfactory aspect of our study is that the correct
quantitative behaviour for $\lx \to 0$ is not reproduced by
the renormalization-group approach. This is apparent in fig. 4, where
the energy difference $E_1-E_0$ is depicted. This quantity approaches
zero linearly with $\lx$, instead of
$\sim  \exp \left[ - 1/\left(3 \sqrt{2}\lx \right) \right]$,
as predicted by the dilute-gas instanton approximation.
We conclude that the first order of the
derivative expansion cannot account quantitatively for the
non-local effects associated with tunnelling. 

However, the use of
an alternative method, such as an expansion around a saddle point,
can remedy the situation. A very good example is provided by the
study of tunnelling rates in three dimensions \cite{first2}.
The renormalization group can be used for the integration of
high frequency fluctuations (that may even induce radiative symmetry
breaking), while an expansion around a saddle point (such as the 
scalar instanton or bounce) can account for tunnelling effects.
The complementarity of the two methods guarantees the
consistency of the calculation.

\paragraph{An approximate analytical solution:} 
As a final confirmation of the previous discussion we present an
approximate analytical solution of the evolution equation near the
origin of the potential. By defining the quantities
$\rho=\phi^2/2$, $\rht=k\rho$, $u_k(\rht)=U_k(\rho)/k$, $t=\log\,k$, 
we can write eq. (\ref{eveq}) as
\be
\frac{\partial u_k'(\rht)}{\partial t} = -2 u_k' -\rht \,u_k'' 
-\frac{1}{2 \pi} \frac{3u_k''+2\rht\,u_k'''}{1+u_k'+2\rht\,u_k''}.
\label{reseveq1} \ee
Primes on $u_k(\rht)$ denote derivatives with respect to $\rht$.
Near the origin of the potential ($\rht \simeq 0$) we can approximate
the above equation by
\be
\frac{\partial u_k'}{\partial t} + \left(
\rht + \frac{3}{2 \pi} \frac{1}{1+u_k'} \right)
\frac{\partial u_k'}{\partial \rht} +2 u_k' =0.
\label{reseveq2} \ee
This is a first-order partial differential equation for $u_k(\rht)$, 
which can be solved with the method of characteristics. 

The most general solution is
\beq
\rht \sqrt{u'_k} + \frac{3}{2 \pi} \tan^{-1} \sqrt{u_k'} 
&=& F \left( u_k' e^{2t} \right)
~~~~~~~~~{\rm for}~~u'_k>0
\nonumber \\
\rht \sqrt{-u'_k} + \frac{3}{4 \pi} \log\left(
\frac{1+\sqrt{-u_k'}}{1-\sqrt{-u_k'}}  \right)
&=& G \left( u_k' e^{2t} \right)
~~~~~~~~~{\rm for}~~u'_k<0.
\label{apsol1} \eeq
The functions $F$ and $G$ are determined by the initial condition for
the potential at $k=\Lx$. For the choice of eq. (\ref{initial})
with $m^2=1$ we find 
\beq
F(x) &=& \frac{1}{8 \lx} \left(1 +x \right)\sqrt{x} + 
\frac{3}{2\pi} \tan^{-1} \left( \frac{\sqrt{x}}{\Lx} \right)
\nonumber \\
G(x) &=&  \frac{1}{8 \lx} \left(1 +x \right)\sqrt{-x} + 
\frac{3}{4\pi} \log \left( 
\frac{1+\frac{\sqrt{-x}}{\Lx}}{1-\frac{\sqrt{-x}}{\Lx}} \right).
\label{apsol2} \eeq

The above solution displays all the characteristic behaviour 
we discussed earlier. The minimum of the potential always moves to
the origin at a scale $k_{cr}=12\lx/\pi$, while the pole at
$k^{-2} \partial^2 U_k/ \partial \phi^2 \simeq u'_k = -1$ is never crossed. 
We can also obtain an approximate expression for the mass term at
the origin $(\rht =0)$ in the limit $k\to 0$, $u'_k(0) \to \infty$. 
We find $\sqrt{\partial U_{k\to 0}(0) /\partial \rho }
= \sqrt{ \partial^2 U_{k \to 0}(0)/ \partial \phi^2}
\simeq 6 \lx$, for small $\lx$. This explains the linear dependence of 
$\Delta E$ on $\lx$, observed in fig. 4 for small $\lx$, even though the
predicted slope is not quantitatively correct.

\paragraph{Acknowledgements:} 
The work of N.T. was supported by the E.C. under TMR contract 
No. ERBFMBICT-983132.

\newpage


\newpage

\noindent
{\Large \bf Figures}

\begin{itemize}

\item 
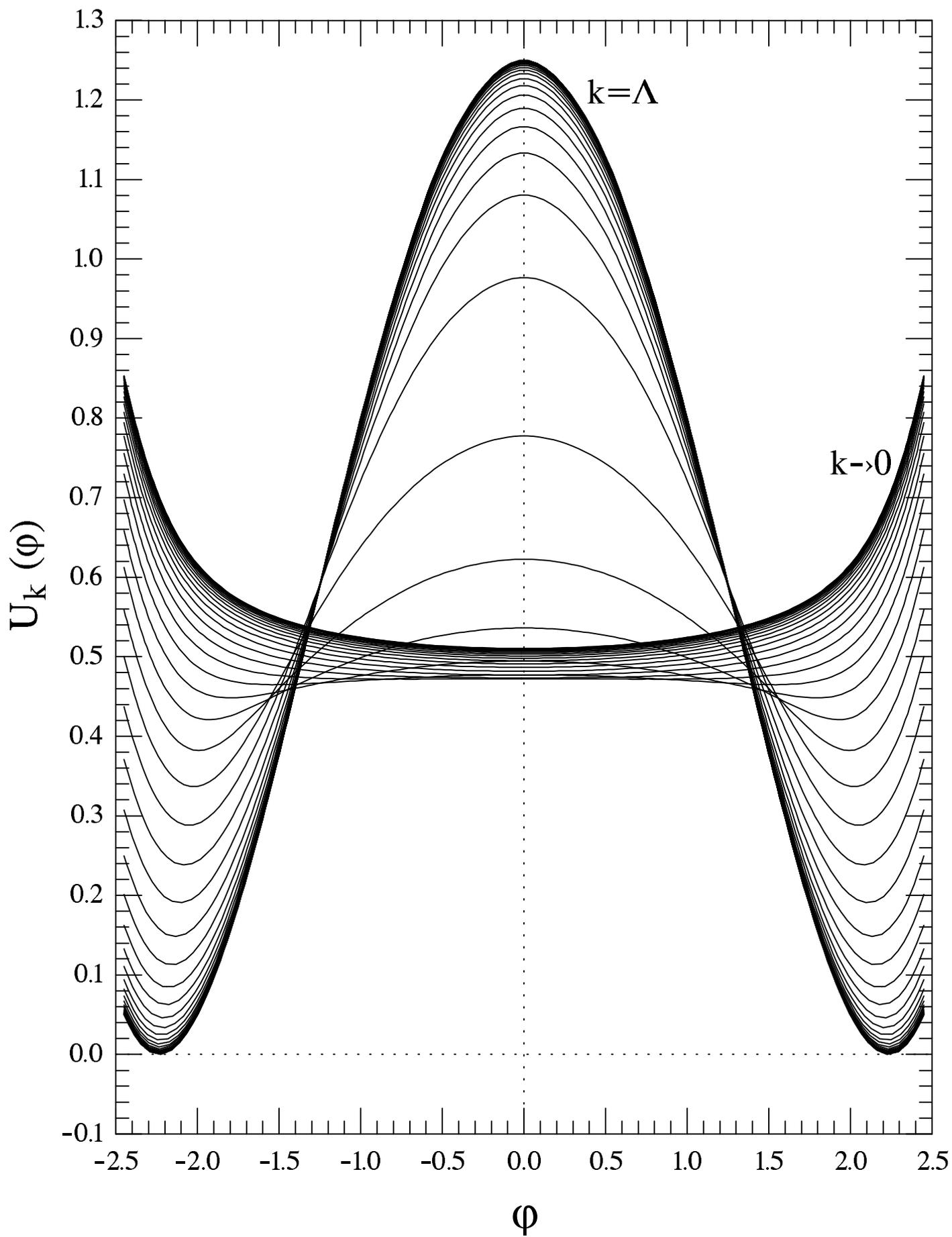
Fig. 1: The evolution of the potential as the coarse-graining scale is
lowered from $k=\Lx \gg 1 $ to $k=0$.
The initial form of the potential is given by eq. (\ref{initial}) 
with $m^2=1$ and $\lx=0.05$.

\item
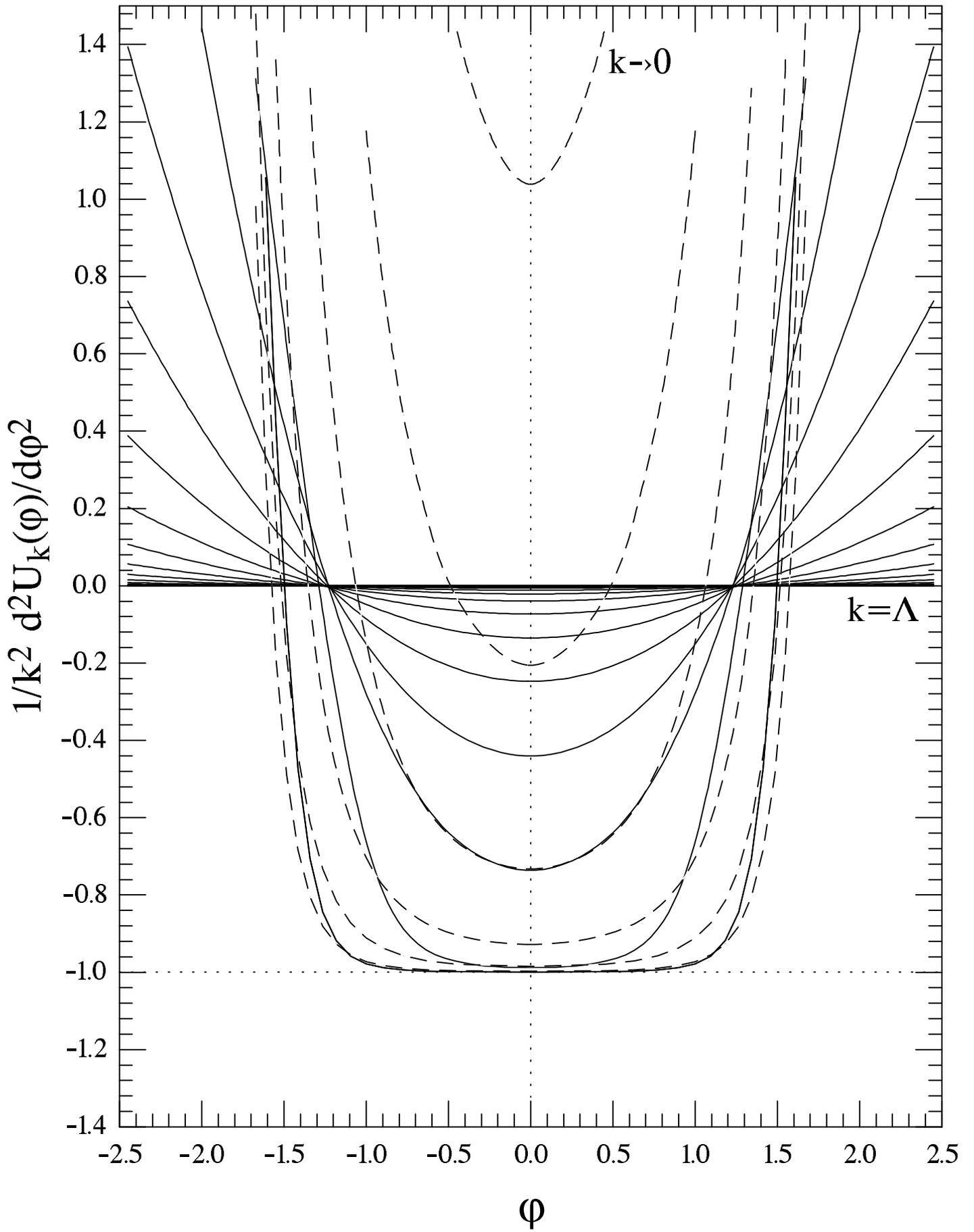
Fig. 2: Same as in fig. 1 for 
$\left[\partial^2 U_k/\partial \phi^2\right]/k^2$. 

\item
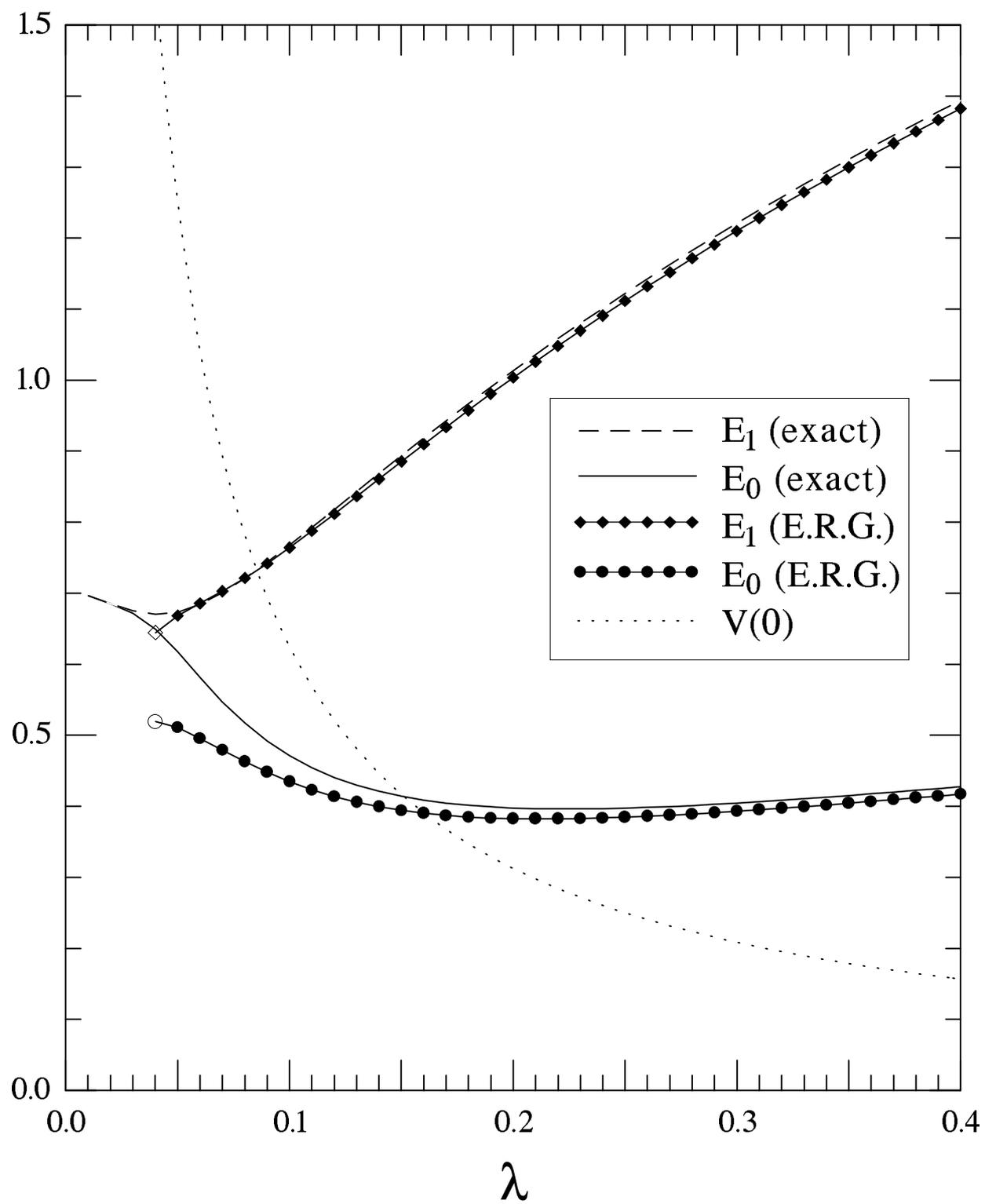
Fig. 3: The first two energy eigenvalues 
$E_0$, $E_1$, 
as predicted by the exact renormalization group in the lowest order of
the derivative expansion, along with the exact results, as a function
of $\lx$, for the potential of eq. (\ref{initial}) with $m^2=1$. 
The dotted line indicates the height of the barrier for the potential of
eq. (\ref{initial}).

\item 
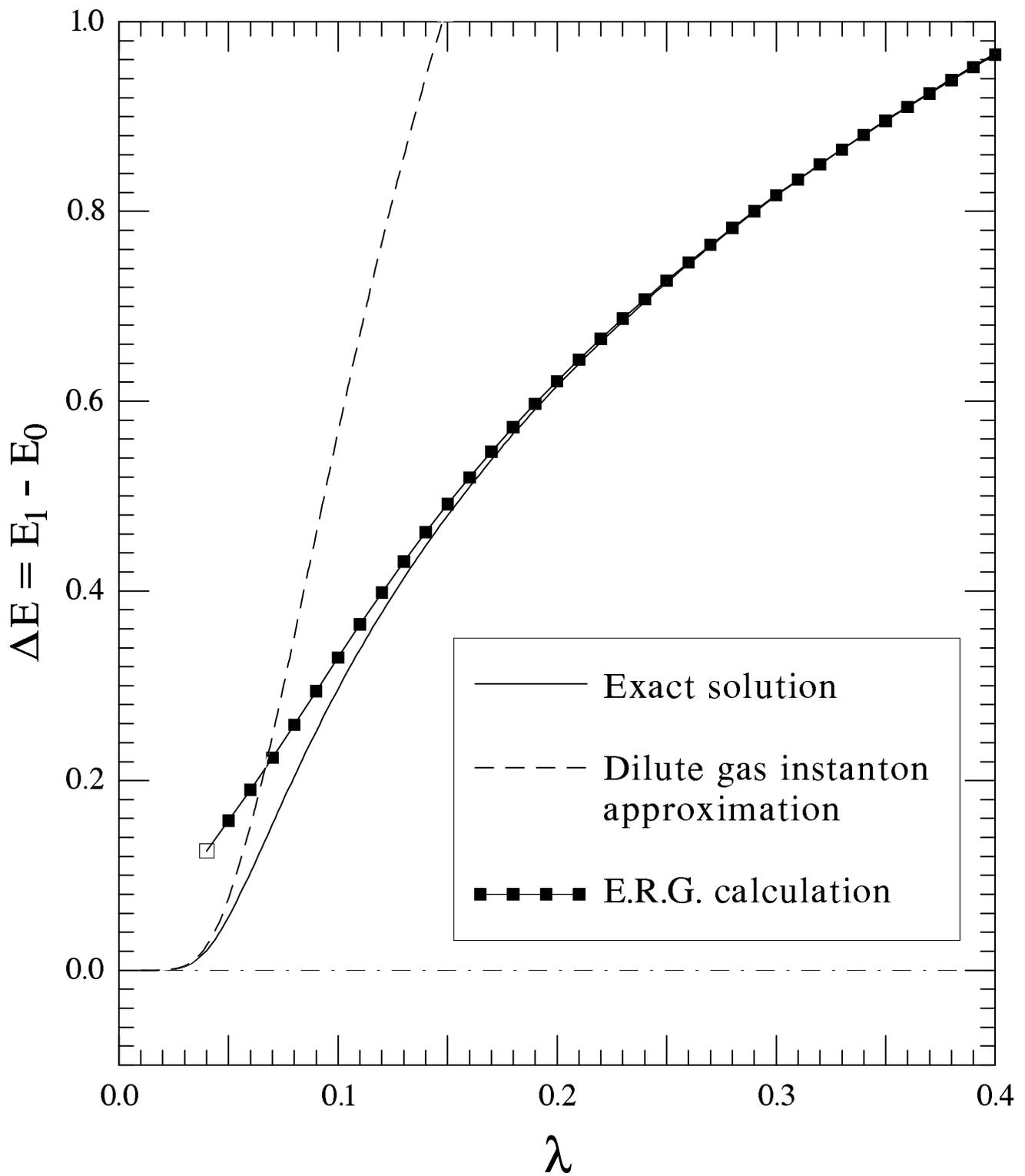
Fig. 4: The energy gap $E_1-E_0$ as a function of $\lx$.
We display the exact values, the predictions of the exact renormalization
group in the lowest order of the derivative expansion, and the 
results of the dilute-gas instanton approximation.

\item
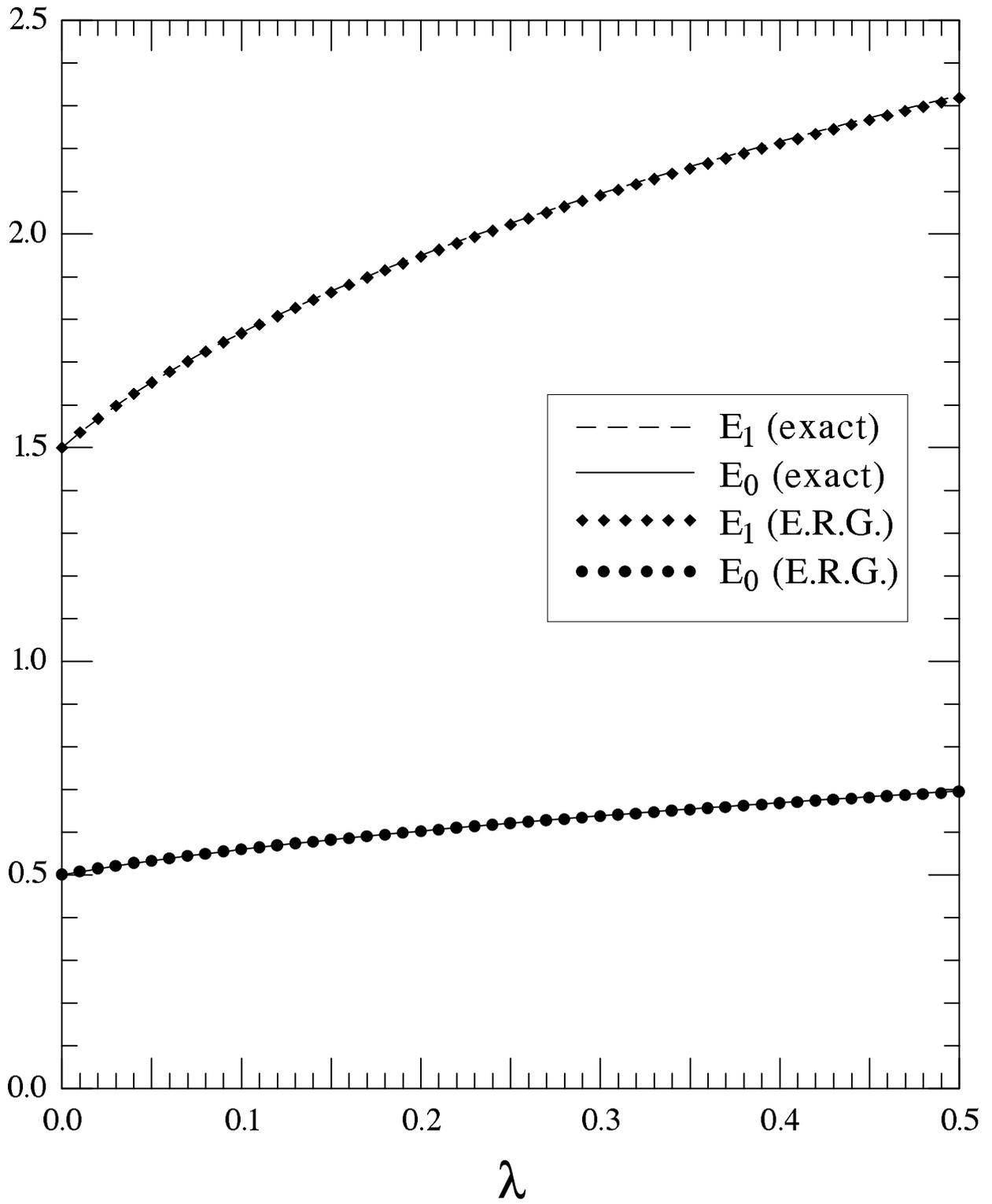
Fig. 5: Same as in fig. 3 for the potential of eq. (\ref{initial})
with $m^2=-1$. 

\end{itemize}


\begin{thebibliography}{nn}


\bibitem{kad} L.P.~Kadanoff, Physica {\bf 2} (1966) 263. 

\bibitem{wil} K.~G.~Wilson, \PR{B4} (1971) 3174; 3184
  K.~G.~Wilson and I.~G.~Kogut, \PRP{12} (1974) 75.

\bibitem{rg} F. Wegner and A. Houghton, \PR{A8} (1973) 401;
  F.J.~Wegner in {\em Phase Transitions and Critical Phenomena},
  vol.~6, eds.~C.~Domb and M.S.~Greene, Academic Press (1976);
J.F.~Nicoll and T.S.~Chang, \PL{A62} (1977) 287;
T. S. Chang, D. D. Vvedensky and J. F. Nicoll, 
Phys. Rep. {\bf 217} (1992) 280;
S. Weinberg, {\em Critical phenomena for field theorists}, 
in {\em Understanding the Fundamental Constituents of Matter}, p. 1, 
Ed.\ by A.\ Zichichi (Plenum Press, N.Y. and London, 1978)
J. Polchinski, \NP{B231} (1984) 269;
A.~Hasenfratz and P.~Hasenfratz, \NP{B270} (1986)
  685; P.~Hasenfratz and J.~Nager, \ZP{C37} (1988) 477;
M. Reuter and  C. Wetterich, \NP{B427}
(1994) 291.

\bibitem{wet} C.~Wetterich, \PL{B301} (1993) 90.

\bibitem{averact} C. Wetterich, \NP{B352} (1991) 529; \ZP{C57} (1993)
  451; {\bf C60} (1993) 461.

\bibitem{review}
J. Berges, N. Tetradis, C. Wetterich, preprint hep-ph/0005122,
to appear in Physics Reports.

\bibitem{indices} N. Tetradis and C. Wetterich, \NP{B422} [FS] (1994)
  541.

\bibitem{morris}  
T.~Morris, \PL{B329} (1994) 241; \PL{B334} (1994) 355;
Int. J. Mod. Phys. {\bf A9} (1994) 2411;
Nucl. Phys. {\bf B458} [FS] (1996) 477;
Nucl. Phys. {\bf B495} (1997) 477; 
Int. J. Mod. Phys. {\bf B12} (1998) 1343;
T. R. Morris and J. F. Tighe, JHEP {\bf 9908} (1999) 007;
T.R.~Morris and M.D.~Turner, Nucl. Phys. {\bf B509} (1998) 637.

\bibitem{more}  
K.-I. Aoki, K. Morikawa, W. Souma, J.-I. Sumi and H. Terao,
Prog. Theor. Phys. {\bf 95} (1996) 409;
Prog. Theor. Phys. {\bf 99} (1998) 451.

\bibitem{first1} J.~Berges and C.~Wetterich, \NP{B487} (1997) 675;
J. Berges, N. Tetradis, C. Wetterich, Phys. Lett. {\bf B393} (1997) 387;
S. Seide and C. Wetterich, Nucl. Phys. {\bf B562} (1999) 524.

\bibitem{first2}
A. Strumia and N. Tetradis, 
\NP {B542} (1999) 719;
A. Strumia, N. Tetradis and C. Wetterich, 
\PL {B467} (1999) 279;
A. Strumia and N. Tetradis, 
Nucl. Phys. {\bf B554} (1999) 697;
Nucl. Phys. {\bf B560} (1999) 482;
JHEP {\bf 9911} (1999) 023.

\bibitem{convex}
N. Tetradis and C. Wetterich, Nucl. Phys. {\bf B383} (1992), 197;
N. Tetradis and D. F. Litim, Nucl. Phys. {\bf B464} [FS] 
(1996) 492.

\bibitem{qmt}
K.-I. Aoki, A. Horikoshi, M. Taniguchi and H. Terao,
in the proceedings of the Workshop on {\em The Exact Renormalization
Group}, Faro, Portugal, September 1998 (World Scientific, Singapore, 1999),
preprint hep-th/9812050.

\bibitem{ramond}
P. Ramond, {\em Field Theory, A Modern Primer} (Benjamin/Cummings, Menlo Park,
California, 1981). 

\bibitem{num}
J.~Adams, J.~Berges, S.~Bornholdt, F.~Freire, N.
  Tetradis and C. Wetterich, \MPL{A10} (1995) 2367.

\bibitem{coleman} 
S.~Coleman, {\em Aspects of Symmetry} (Cambridge University Press, Cambridge, 
1985).

\end{thebibliography}
\end{document}